\documentclass[fleqn,twoside]{article}
\usepackage{espcrc2}
\usepackage{psfig}

\def\mod{\mathop{\rm mod}}

\begin{document}

\title{
{\vspace{-1.2em} \parbox{\hsize}{\hbox to \hsize 
{\hss  \normalsize IFUP-TH 2002/31, TRINLAT-02/03}}} \\
A Hamiltonian lattice study of the two-dimensional Wess-Zumino model}

\author{
Matteo Beccaria\address{Dipartimento di Fisica
dell'Universit\`a di Lecce and I.N.F.N., Sezione di Lecce},
Massimo Campostrini\address{Dipartimento di Fisica ``Enrico Fermi''
dell'Universit\`a di Pisa and I.N.F.N., Sezione di Pisa}%
\thanks{Presented by M.\ Campostrini} and
Alessandra Feo\address{School of Mathematics,
Trinity College, Dublin 2}
}

\begin{abstract}
We investigate a Hamiltonian lattice version of the two-dimensional
Wess-Zumino model by Quantum Monte Carlo simulations.  In order to
study the pattern of supersymmetry breaking, we measure the ground
state energy and the correlation length along a trajectory approaching
the continuum limit.  The algorithm is very effective in measuring the
ground state energy, and adequate for the correlation length.
\end{abstract}

\maketitle

\section{INTRODUCTION}

Numerical simulations of lattice field theories are usually performed
in the Lagrangian formulation.  
Nonetheless, we think there are very good reasons to
develop numerical simulation techniques for the Hamiltonian approach~\cite{KS}:
powerful many-body techniques are available~\cite{QMC}, which allow
the direct computation of the vacuum wave function properties;
fermions are implemented directly and need not be integrated out;
properties like the mass spectrum are more immediate.
Finally, universality checks between the Lagrangian and the
Hamiltonian formalism are very welcome.

\section{THE MODEL}
We study the Hamiltonian lattice version of the two-dimensional
Wess-Zumino model described in Refs.~\cite{lattice2001,Trento}; we
only wish to highlight here the main features of the formulation.

In the Hamiltonian formalism, since $H$ is conserved, it is possible to
preserve exactly a 1-dimensional subalgebra of the original supersymmetry
algebra, i.e., we can write $H=Q^2$, where $Q$ is a fermionic charge.
This subalgebra is enough to guarantee some of the most important
property of supersymmetry, including a non-negative spectrum, 
and pairing of fermionic and bosonic
states of nonzero energy; spontaneous breaking of supersymmetry is
equivalent to a strictly positive ground-state energy $E_0$; the full
supersymmetry algebra is recovered in the continuum limit together
with Lorentz invariance.

In order to obtain a Hamiltonian free of fermion sign problems, and
therefore amenable to Quantum Monte Carlo methods, we adopt free boundary
conditions, with lattice size $L \equiv 2\;(\mod 4)$.

The model is parametrized by a {\em prepotential\/} $V(\phi)$,
an arbitrary polynomial in the bosonic field.
The two-dimensional Wess-Zumino model is superrenormalizable; 
fields do not renormalize, and only $V(\phi)$ needs to be
normal ordered.

In strong coupling at leading order, the model reduces to independent
copies of supersymmetric quantum mechanics, one for each site;
supersymmetry is broken if and only if the degree of the prepotential
$V$ is even~\cite{Witten1}.  In weak coupling, on the other hand,
supersymmetry is broken at tree level if and only if $V$ has no
zeroes.  The predictions of strong coupling and weak coupling are
quite different, and it is interesting to study the crossover from
strong to weak coupling.

\section{MONTE CARLO SIMULATIONS}

We perform our simulations using the Green Function Monte Carlo (GFMC)
algorithm~\cite{QMC}.  A discussion of GFMC in the context of the
present problem can be found in Ref.~\cite{Trento}; we only wish to
remark the main features of the algorithm: the aim is to generate a
stochastic representation of the ground-state wavefunction, which is
then used to compute expectation values of observables.  Statistical
fluctuations are reduced with the help of a guiding wavefunction,
whose free parameters are determined dynamically during the
simulation. 
In order to keep the variance of observables finite as the simulation
proceeds, it is necessary to simulate a population of $K$ {\em
walkers\/} (field configurations at fixed time), and extrapolate the
results to $K\to\infty$.

We focus on the case $V = \lambda_2\phi^2 + \lambda_0\phi$; strong
coupling always predicts supersymmetry breaking; weak coupling
predicts unbroken supersymmetry for $\lambda_0<0$; according to Ref.\ 
\cite{Witten2}, unbroken supersymmetry should be accompanied by a
nonzero $\langle\phi\rangle$ (parity breaking).

We presented preliminary results, obtained by GFMC, for intermediate
couplings in Ref.~\cite{lattice2001}; our aim is to extend the study
towards the continuum limit and to larger lattices.

Perturbative computations show that
\begin{equation}
\lambda_2 \approx a \lambda_2^{\rm ren}, \quad
\lambda_0 \approx a \lambda_0^{\rm ren} +
a \lambda_2^{\rm ren}\frac{1}{2\pi}\ln\left(aM\right),
\label{eq:evol2_L}
\end{equation}
where $\lambda_i$ is the adimensional lattice bare coupling,
$\lambda_i^{\rm ren}$ is the renormalized (continuum) coupling, with
dimension of $m^1$, defined at the mass scale $M$, and $a$ is the
lattice spacing.
We study, as $\lambda_2\to 0$, the trajectory
\begin{equation}
\lambda_0={\lambda_2\over2\pi}\,\ln(4\lambda_2),
\label{eq:trajectory}
\end{equation}
corresponding to a perturbative RG trajectory (\ref{eq:evol2_L}); the
effect of $\lambda_0$ is small in the range we considered,
therefore we expect Eq.\ (\ref{eq:trajectory}) to be a reasonable
approximation to a true RG trajectory.

We estimate the correlation length from the exponential decay of the
connected correlation function $G_d=\langle\phi_n\phi_{m}\rangle_c$
averaged over all $n,m$ pairs with $|m-n|=d$, excluding pairs for
which $m$ or $n$ is closer to the border than (typically) 8.
In our formulation, fermions are staggered and 
even/odd $d$ correspond to different channels.

\begin{figure}[tb]
\null\vskip2mm
\centerline{\psfig{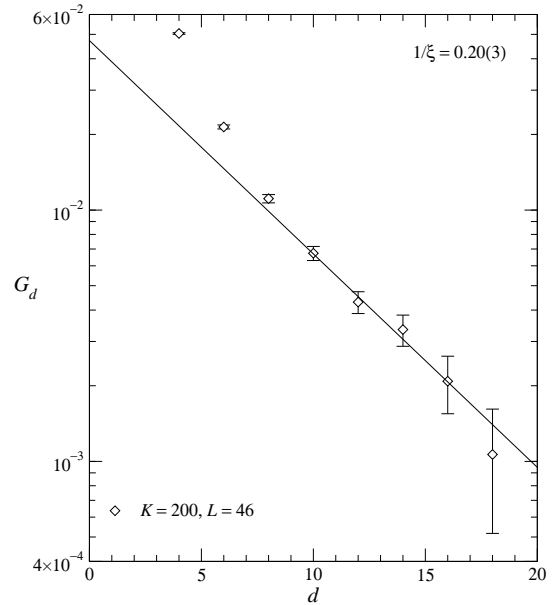}}
\vskip-8mm
\caption{Connected correlation of $\phi$ at even distance for
$V=0.35\,\phi^2+0.02$; the solid line is an exponential fit from
distance 8 to 18; $L=34$ and $K=100$ data are consistent with the
data shown.}
\label{fig:xieven,l2=0.35}
\end{figure}

\begin{figure}[tb]
\null\vskip2mm
\centerline{\psfig{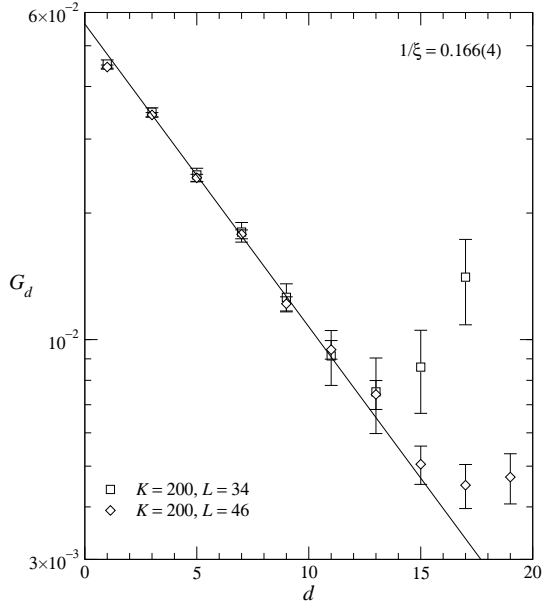}}
\vskip-8mm
\caption{Connected correlation of $\phi$ at odd distance for
$V=0.35\,\phi^2+0.02$; the solid line is an exponential fit from
distance 3 to 15; $K=100$ data are consistent with the $K=200$ data shown.}
\label{fig:xiodd,l2=0.35}
\end{figure}

We begin with the discussion of the case $V=0.35\,\phi^2+0.02$, for which we have obtained the
statistics of $4{\times}10^6$ GFMC iterations.  The even-$d$ channel
is plotted in Fig.\ \ref{fig:xieven,l2=0.35}; it is very difficult to
extract a correlation length, presumably because $\phi$ has a very
small overlap with the lightest state of the channel, and the value of
$\xi$ quoted in Fig.\ \ref{fig:xieven,l2=0.35} should be considered
tentative.  The odd-$d$ channel, plotted in Fig.\ 
\ref{fig:xiodd,l2=0.35}, is much cleaner, and it is possible to
estimate $\xi$ with a good precision.

\begin{figure}[tb]
\null\vskip2mm
\centerline{\psfig{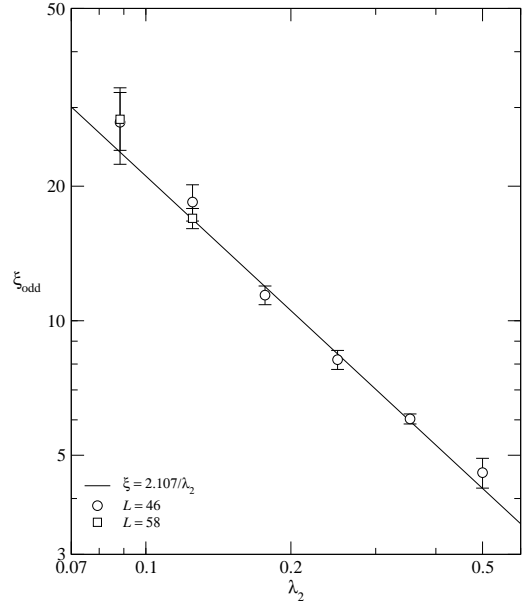}}
\vskip-8mm
\caption{Correlation length in the odd distance channel
vs.\ $\lambda_2$; the solid line is a scaling fit.}
\label{fig:xioddlog}
\end{figure}

For the other values of $\lambda_2$, the situation is similar but with
larger errors; we have a statistics of at least $10^6$ iterations,
which we are increasing to $4{\times}10^6$.  The values of $\xi_{\rm
odd}$ follow nicely the expected behavior $\xi\propto1/\lambda_2,$ as
shown in Fig.\ \ref{fig:xioddlog}: the entire range
$0.088\le\lambda_2\le0.35$ seem to be in the scaling region, with
$\lambda_2=0.5$ a borderline case.  The values of $\xi_{\rm even}$
have very large errors, and it is hard to draw any conclusion from
them.

We measure the ground state energy $E_0$ along the trajectory
(\ref{eq:trajectory}); the measurements have a very small statistical
error, ranging from 1\% for $\lambda_2=0.044$ (where
$E_0/L\approx10^{-3}$) to 0.1\% for $\lambda_2=0.5$.  We
extrapolate to $L\to\infty$ and $K\to\infty$ fitting $E_0/L$ to the
form
\begin{equation}
{E_0/L} = {\cal E}_0\left(1 + {c_L/L} + {c_K/K}\right).
\label{eq:E0/L}
\end{equation}

\begin{figure}[tb]
\null\vskip2mm
\centerline{\psfig{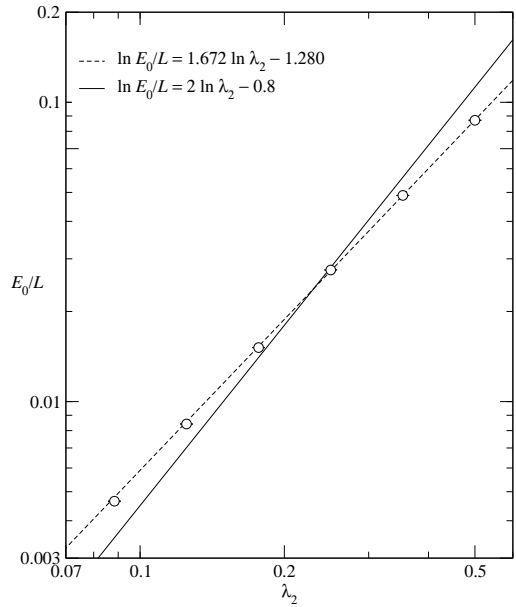}}
\vskip-8mm
\caption{Ground-state energy density vs.\ $\lambda_2$.  The dashed
line is the result of a fit; the solid line shows the na\"\i ve
scaling behavior.}
\label{fig:E0log}
\end{figure}

$E_0/L$ is plotted in Fig.\ \ref{fig:E0log}:
it seems to behave $\propto\lambda_2^{5/3}$, while na\"\i ve
scaling would predict $\propto\lambda_2^2$.  The value of $E_0/L$
(disregarding this puzzling exponent) and the lack of any signal for a
breakdown of parity (like a double-peaked distribution of $\phi$)
strongly hint that the trajectory (\ref{eq:trajectory}) belongs to the
phase with broken supersymmetry and zero $\langle\phi\rangle$.
We are repeating the computation for trajectories with smaller
$\lambda_0$.

\end{document}